\documentclass[aip,rsi,reprint,graphicx,floatfix]{revtex4-1}
\pdfoutput=1

\usepackage{amsmath}
\usepackage{graphicx}
\usepackage{epstopdf}
\usepackage{natbib}
\usepackage{verbatim}
\usepackage[withbib, all]{authorindex}	
\usepackage[usenames]{color}
\usepackage{bm}
\usepackage{siunitx}
\DeclareSIUnit\torr{torr}
\DeclareSIUnit\gauss{G}

\begin{document}

\title{A versatile LabVIEW and FPGA-based scanned probe microscope for in-operando electronic device characterization}

\author{Andrew J. Berger}
\email{berger.156@osu.edu}
\affiliation{Department of Physics, The Ohio State University, Columbus, Ohio 43210, USA}

\author{Michael R. Page}
\affiliation{Department of Physics, The Ohio State University, Columbus, Ohio 43210, USA}

\author{Jan Jacob}
\affiliation{Werum Software \& Systems CIS AG, Wulf-Werum-Stra{\ss}e 3, 21337 L{\"u}neburg, Germany}

\author{Justin R. Young}
\affiliation{Department of Physics, The Ohio State University, Columbus, Ohio 43210, USA}

\author{Jim Lewis}
\affiliation{National Instruments, Austin, Texas 78759, USA}

\author{Lothar Wenzel}
\affiliation{National Instruments, Austin, Texas 78759, USA}

\author{Vidya P. Bhallamudi}
\affiliation{Department of Physics, The Ohio State University, Columbus, Ohio 43210, USA}

\author{Ezekiel Johnston-Halperin}
\affiliation{Department of Physics, The Ohio State University, Columbus, Ohio 43210, USA}

\author{Denis V. Pelekhov}
\affiliation{Department of Physics, The Ohio State University, Columbus, Ohio 43210, USA}

\author{P. Chris Hammel}
\affiliation{Department of Physics, The Ohio State University, Columbus, Ohio 43210, USA}

\date{\today}

\begin{abstract}
Understanding the complex properties of electronic and spintronic devices at the micro- and nano-scale is a topic of intense current interest as it becomes increasingly important for scientific progress and technological applications.  In-operando characterization of such devices by scanned probe techniques is particularly well-suited for the microscopic study of these properties.  We have developed a scanned probe microscope (SPM) which is capable of both standard force imaging (atomic, magnetic, electrostatic) and simultaneous electrical transport measurements.  We utilize flexible and inexpensive FPGA (field programmable gate array) hardware and a custom software framework developed in National Instrument's LabVIEW environment to perform the various aspects of microscope operation and  device measurement.  The FPGA-based approach enables sensitive, real-time cantilever frequency-shift detection.  Using this system, we demonstrate electrostatic force microscopy of an electrically-biased graphene FET device.  The combination of SPM and electrical transport also enables imaging of the transport response to a localized perturbation provided by the scanned cantilever tip.  Facilitated by the broad presence of LabVIEW in the experimental sciences and the openness of our software solution, our system permits a wide variety of combined scanning and transport measurements by providing standardized interfaces and flexible access to all aspects of a measurement (input and output signals, and processed data).  Our system also enables precise control of timing (synchronization of scanning and transport operations) and implementation of sophisticated feedback protocols, and thus should be broadly interesting and useful to practitioners in the field.  
\end{abstract}

\maketitle

\section{Introduction}
Scanned probe imaging is a versatile tool for studying, with high spatial resolution, many interesting physical phenomena (magnetism, surface roughness, conductivity, etc.).  The toolkit of scanned probe techniques has been expanded by the use of a scanned proximal probe for imaging and local perturbation in conjunction with simultaneous electrical transport measurements.  For example, quantized conductance and universal conductance fluctuations have been mapped by scanned gate imaging \cite{topinka_imaging_2000, berezovsky_imaging_2010}, Kelvin probe microscopy has been used to characterize charge traps in AlGaN/GaN HEMTs \cite{cardwell_nm-scale_2012}, and a Hall cross was used to quantify and map the magnetic field of an MFM cantilever \cite{panchal_magnetic_2013}.  One goal of this article is to provide a guide to the fundamental steps necessary to reproduce an SPM with such capabilities.  Commercially-available SPM systems facilitate simple and quick sample analysis, providing high resolution and scan rates, but are typically focused on measurements of passive sample properties (topography, e.g.).  In order to expand these capabilities to monitor active devices, the measurement sequence must interface with external hardware and allow end-user modification for flexible measurement protocols.   Synchronization of imaging and acquisition of transport data is of particular importance.  In addition to electrical connectivity for devices, some varieties of transport measurements also require specific environments and functionalities, such as magnetic field application, vacuum, and cryogenic compatibility.  Here we describe an integrated, low cost, custom-built system that combines all of these capabilities in an instrument that provides both conventional scanning probe force microscopy and mapping of electrical transport properties as a function of probe position.   

The primary challenges for in-operando SPM are the precise positioning of the scanned probe relative to the device, connecting and managing electrical connections to a sample, and flexibility and precision for combining and synchronizing transport and scanned probe measurements and manipulations.  To further complicate matters, these tasks may need to be accomplished within the constraints imposed by magnetic field application (the confined space of an electromagnet), in vacuum, and at cryogenic temperatures.  Proper cantilever-sample positioning requires: accurately locating and positioning the active area of the device, whose size is often small relative to the total substrate area, determining and controlling scan height, and minimizing vibrational noise and drift (thermal or magnetic field-influenced).  Managing electrical connectivity involves enabling relatively easy and reconfigurable wiring of samples in a geometry that provides cantilever access to the device, electrostatic-discharge protection of sensitive devices, and minimization of electrical crosstalk and noise.  To permit flexibility of the measurement protocol, an open hardware/software architecture that can be easily modified by the user is necessary.  In this article, we present our solutions to these challenges.

A broadly useful system will allow the operator easy access and flexibility with regards to both hardware connections and software processing.  This was the primary factor in our selection of the FPGA and LabVIEW-based microscope control solution.  The FPGA (Field-Pro\-gram\-ma\-ble Gate Array) provides user-defined reprogrammable processing, while ensuring the fast, deterministic control necessary for scanned probes.  Our system relies on open interfaces---industry standard BNC connections and GPIB communications---which allow easy expansion and flexible combination with standard measurement equipment.  In-operando imaging could use a variety of inputs as the imaging parameter: cantilever frequency or amplitude, device conductance, Hall voltage, cantilever tip bias (as in Kelvin probe microscopy \cite{girard_electrostatic_2001}), and others.  Our hardware/software framework makes it easy to add imaging parameters due to its reprogrammability and LabVIEW's extensive library of third-party device drivers.  For example, incorporation of dedicated current and voltage sources and meters is straightforward with GPIB communication.  The microscope user can then select the most useful imaging parameter or collect multiple parameters simultaneously.  

The accessibility and edit-ability of the software, written in the widely-used LabVIEW programming language, offers the user precise control of measurement sequencing.  Scanned probe operations can be synchronized with transport measurements, scan parameters (e.g. tip bias, or scan height) can be tuned in response to transport, and actions or procedures can be easily added to the measurement protocol.  This is a shift of focus relative to commercial SPM solutions, which are typically optimized for ease-of-use and provide less software and hardware re-configurability for customized measurement protocols.  

Lastly, while home-built SPM controllers often require several custom analog circuits (PID controllers, cantilever phase/frequency detection circuitry), our FPGA/LabVIEW architecture is a complete microscope control solution, requiring no additional home-built circuitry.  This also permits much more versatility than hardwired circuits, which may require rebuilding if different measurement parameters are required (for example, interchanging cantilevers with different resonant frequencies).  This combination of features and processing tools ensures the flexibility and performance to implement and execute a large variety of measurements.
  
\section{Experimental Setup}

\begin{figure*}[htb]
	\centering
		\includegraphics[width=\linewidth]{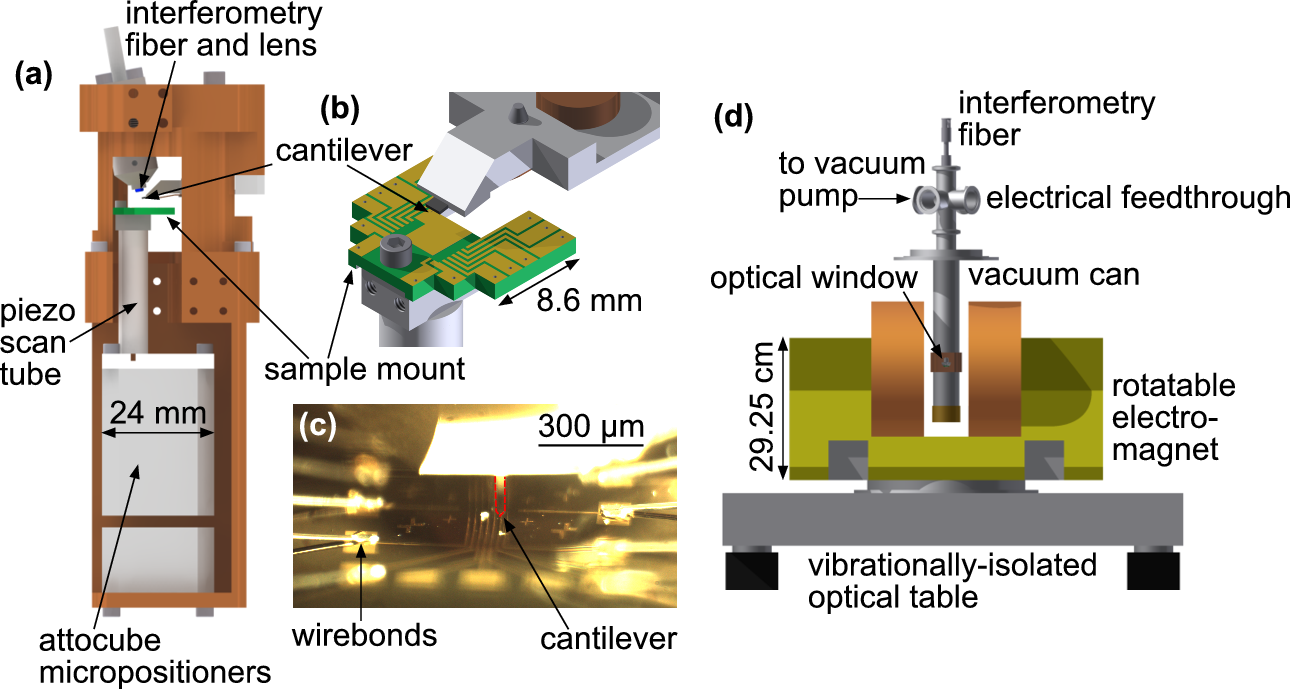}
	\caption{(a) Microscope scan head. (b) Isometric view of cantilever and electrical sample holder. (c) View from optical window of sample-cantilever alignment. Both lithographic and wirebond leads to device are visible.  The cantilever, outlined in red, is also visible.  (d) View of entire experiment, including optical table and vibrationally-isolated rotatable electromagnet.} 
	\label{fig:MicroscopeDesign}
\end{figure*}

Figure \ref{fig:MicroscopeDesign} shows various levels of detail of the microscope design.  The probe head incorporates a cantilever mounted $10^{\circ}$ from horizontal, fiber interferometric detector for cantilever position sensing, and sample mount with electrical contacts.  The sample mount (a patterned printed circuit board) sits atop the scanning hardware---a piezo tube for fine scanning and attocube \cite{attocube} micro-positioners for coarse motion.  The entire scan head is mounted on and enclosed by gold-plated, oxygen-free high thermal conductivity copper (to enable cryogenic operation;  the gold prevents oxidation of the copper surface).  All components of the SPM are made of non-magnetic materials for operation in an external magnetic field (see Fig. \ref{fig:HardDriveMFM}(b)).  The scan head measurements were chosen to fit inside a standard electromagnet (maximum pole spacing = 3 in).

At its widest, the scan head measures 1.72 in (43.69 mm) in diameter in order to fit inside the 2 in outer diameter (1.87 in inner diameter) vacuum can.  Operation in vacuum reduces damping of the oscillating cantilever, provides a clean environment (especially for surface-sensitive graphene samples), and slows the oxidation of ferromagnetic contacts.  The space constraints imposed by the electromagnet and vacuum can however make cantilever-sample alignment more challenging.  We therefore use a vacuum can with an optical window.  An example of the area viewable through this window is shown in Fig. \ref{fig:MicroscopeDesign}(c).  Using a CMOS camera mounted on a Nikon long-working distance optical microscope (SMZ1500), we are able to focus on and monitor the cantilever position relative to the sample during coarse positioning tasks.  It is critical to achieve good initial coarse alignment to a device.  Acquiring and piecing together many fine scan AFM images in order to locate a device can be very time consuming, especially if the initial scans do not contain obvious device features.

\subsection{Instrumentation}
\subsubsection{Electronics}
Central to the microscope operation is a National Instruments \cite{nationalInstruments} PXI-7851R FPGA and data acquisition card (DAQ).  At its front end, this card is responsible for all analog-to-digital (A/D) conversions of the various input signals and digital-to-analog (D/A) conversions of output control signals (see Table \ref{tab:FPGAIO}).  The A/D inputs and D/A outputs (8 of each, $\pm$\SI{10}{\volt} range) are digitized with 16-bit resolution, providing \SI{305}{\micro\volt} precision.  The inputs are sampled at \SI{750}{\kilo\hertz}; outputs at \SI{1}{\mega\hertz}.  In addition to voltage I/O, the FPGA provides high speed signal processing, with a primary clock rate of \SI{40}{\mega\hertz}.  

An FPGA is a reprogrammable logic chip.  Unlike a CPU, the logic executed by an FPGA is not determined on-the-fly by software.  Instead, the desired code is downloaded to the chip, internally rewiring it in order to construct a hardware implementation of the specific logic.  This enables the code to execute deterministically and reliably, with no added latency due to variable resource demand (as in software processes executed by a PC CPU).  Furthermore, multiple processes can be executed truly in parallel by the FPGA: each process has dedicated circuitry associated with it, and so adjacent processes have no bearing on one another's execution.  The FPGA card is housed in a PXIe-1073 chassis, which communicates via a high speed bridge (MXI data link, 250 MB/s) to the host PC's PCI bus.  The PXI system was chosen to allow system expandability, as it has slots for four additional DAQ or bus interface cards.  Furthermore a dedicated chassis for DAQ cards provides cleaner power and reduces electromagnetic noise as compared to a PCI solution housed in the host PC. To access the voltage I/O of the FPGA, the card connects to an SCB-68 screw terminal connector block via a shielded 68-pin SCSI cable.  We installed the terminal block into a custom breakout box which connects the screw terminals to BNC-style panel-mount connectors, enabling easier connectivity.  Table \ref{tab:FPGAIO} describes all of the FPGA inputs and outputs.

\noindent
\begin{table*}[htp]
\caption{FPGA Inputs and Outputs}
\begin{tabular}{ | l | l | }
		\hline
		\textbf{\textit{FPGA Inputs}} & \textbf{\textit{FPGA Outputs}} \\
		\hline
		AI0 - AC-coupled cantilever interferometer signal & AO0 - self-excitation signal \\
		AI1 - DC-coupled cantilever interferometer signal & AO1 - Piezo Tube +X \\
		AI2 - available & AO2 - Piezo Tube -X \\
		AI3 - available & AO3 - Piezo Tube +Y \\
		AI4 - available & AO4 - Piezo Tube -Y \\
		AI5 - available & AO5 - Piezo Tube Z \\
		AI6 - available & AO6 - Electromagnet control voltage \\		
		AI7 - available & AO7 - available \\
		 \hline	
\end{tabular}
\label{tab:FPGAIO}
\end{table*}

\subsubsection{Positioning}
For coarse positioning, we use attocube piezo stepper stages ANPx101 (for x and y motion) and ANPz101, each with 5 mm of travel.  An attocube ANC-150 controller produces the voltage pulses to move the stages.  The piezo tube which we use for fine scanning (EBL \#4, length = 1 in, diameter = 0.25 in, wall thickness = 0.02 in) has a lateral range of \SI{30}{\micro\meter} and a vertical range of $\sim$\SI{3}{\micro\meter} at room temperature, given the $\pm$\SI{500}{\volt} range of our high voltage amplifiers (Trek \cite{trek} 601C).  The FPGA card outputs control voltages of $\pm$\SI{10}{\volt} for each of the 5 electrodes of the piezo tube ($\pm$x, $\pm$y, z).  The parallelism of the FPGA enables simultaneous control of all piezo tube axes, and the step size resolution (given the 16-bit digital resolution of the D/A) is \SI{0.4}{\nano\meter} in x,y and \SI{0.05}{\nano\meter} in z.  Each of the 5 channels is amplified by a high voltage amplifier (gain = 50) in order to bias the piezo tube appropriately.  Note that a simple op-amp inverter circuit (unity gain) could be used to generate the control voltages for the $-x$ and $-y$ quadrants of the piezo tube if additional FPGA analog outputs are required.

\begin{figure}[htp]
	\centering
		\includegraphics[width=\linewidth]{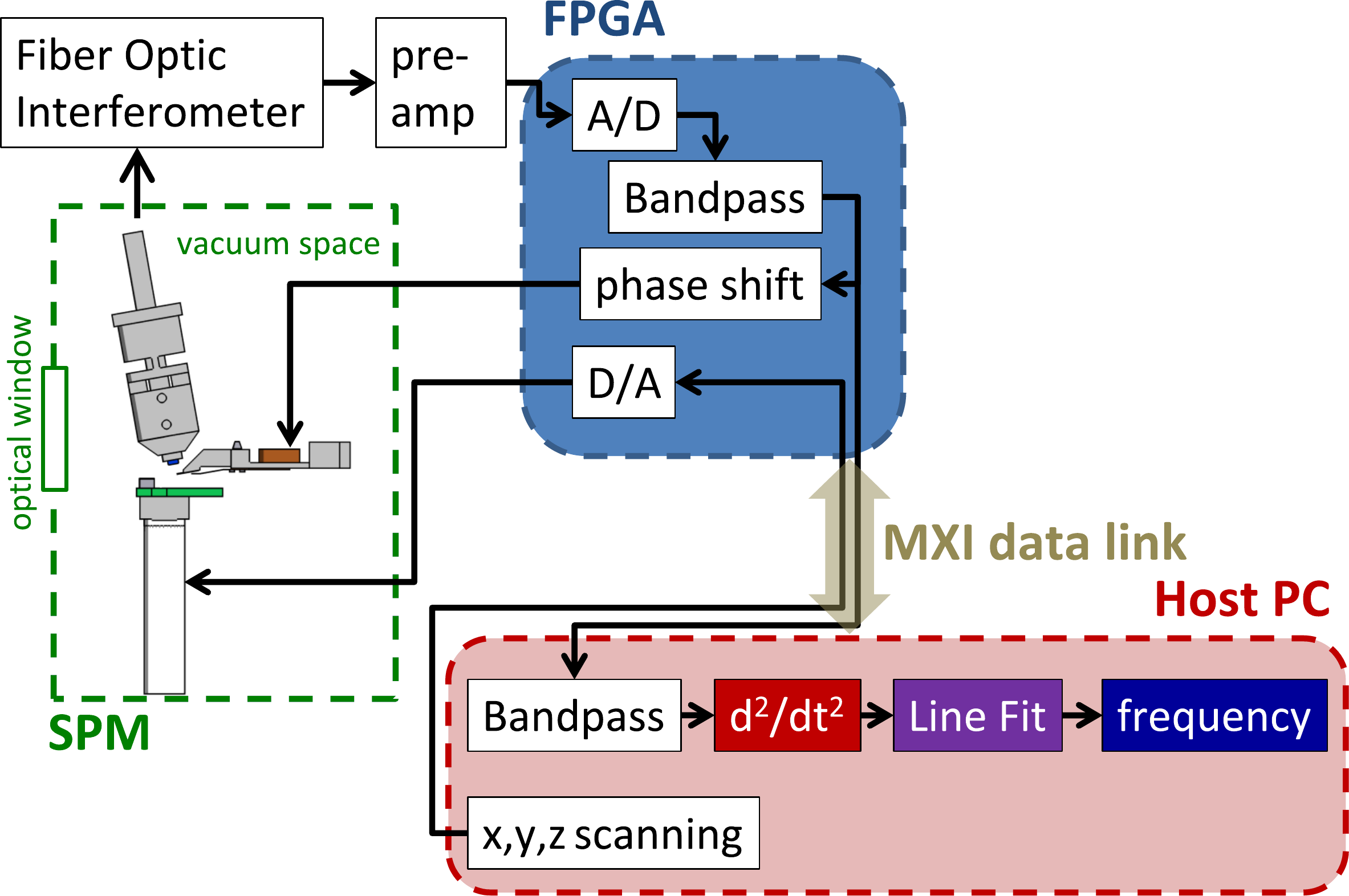}
	\caption{Block diagram of the microscope and control system. The cantilever position signal from the fiber optic interferometer is fed to the A/D converter of the FPGA.  After bandpass filtering, this signal is sent both to the host for further processing (for explanation of frequency detection algorithm, see Sec. \ref{sec:FreqDet} and reference \citenum{obukhov_real_2007}) and to a phase shift function before subsequent output for cantilever self-excitation.  The host PC also calculates scanning voltages, which are sent to the FPGA for D/A conversion and output.} 
	\label{fig:MicroscopeBlockDiag}
\end{figure}

\subsubsection{Cantilever Measurement and Control}
We use a fiber-optic interferometer \cite{rugar_force_1988} to measure the cantilever displacement as a function of time.  Such a system has several implementation advantages for low temperature operation (not demonstrated here) and for use in a confined vacuum space.  However, any system that measures the cantilever position versus time could be substituted for the interferometry system (four-quadrant position-sensing detector, piezoresistive cantilever, etc.), so long as a voltage signal (representing cantilever position) can be provided to the frequency detection software via the DAQ.

The interferometer photoreceiver voltage is fed to the A/D input of the FPGA, which digitizes at a maximum rate of \SI{750}{\kilo\hertz}.  We know by the Nyquist-Shannon theorem \cite{shannon_communication_1949} that the maximum detectable cantilever frequency should therefore be \SI{375}{\kilo\Hz}.  In practice, sampling $\sim$10 times faster than the cantilever frequency of $\sim$\SI{75}{\kilo\Hz} is preferred to achieve our desired frequency measurement precision.  Additionally, because the input range of the A/D is fixed, it is useful to adjust the amplitude of the cantilever photodiode signal (using a pre-amplifier, e.g. Stanford Research \cite{stanfordResearch} SR560) to maximize use of the $\pm$\SI{10}{\volt} range.

\begin{figure}[htp]
	\centering
		\includegraphics[width=\linewidth]{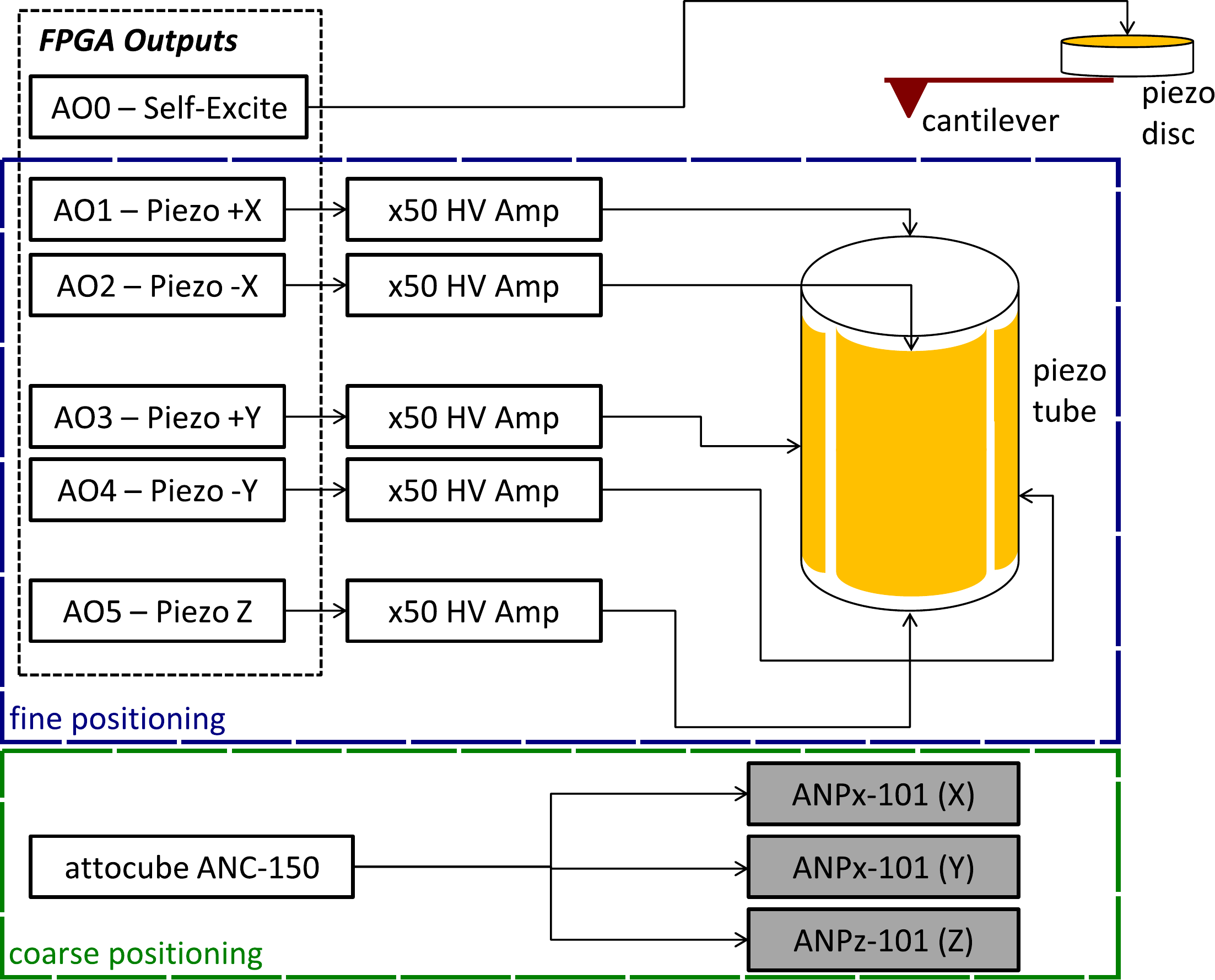}
	\caption{Hardware and electronics for cantilever self-excitation, and sample positioning and scanning.} 
	\label{fig:ScanningHardware}
\end{figure}

While collecting cantilever position data, the FPGA simultaneously provides a periodic voltage to excite the cantilever at its self-determined resonance (see Sec. \ref{sec:SelfExcite}).  This voltage ($\sim$200 mV) is applied to a piezo disc (EBL \#4; 0.25'' diameter, 0.08'' thick) which is mechanically coupled to the cantilever.  Figure \ref{fig:ScanningHardware} shows the various electrical connections for cantilever self-excitation and sample scanning. 

The accessibility of the microscope signal inputs and outputs (photodiode signal, self-excitation drive signal, etc.) affords straightforward interfacing with external hardware.  For example, the user can easily connect the voltage output of the interferometer to a spectrum analyzer and oscilloscope.  Alternatively, the LabVIEW environment makes software implementation of such functionality simple.  

\subsubsection{Transport Measurements and Sample Wiring}
For electrical measurements of transport devices, we use a Keithley \cite{keithley} 6221 AC/DC current source and 2425 DC SourceMeter.  A Stanford Research SR850 lock-in amplifier provides sensitive lock-in detection.  These instruments interface with the PC via GPIB-USB.  Depending on the user's preference for stand-alone measurement hardware versus software-based implementation, these instruments could be implemented by appropriate PXI cards and LabVIEW-based software algorithms, such that the entire microscope could be run from a single chassis and host-PC, providing a compact, relatively inexpensive control system.  Table \ref{tab:FPGAIO} shows that there are several remaining inputs and outputs available which could also be utilized for these purposes.   

Transport measurements of a FET device require reconfigurable electrical wiring (choice of source, drain, gate, voltage probes, etc.).  To this end, we have developed a compact sample stage which allows 16 electrical contacts be made to a device without obstructing cantilever access (Fig. \ref{fig:MicroscopeDesign}(b)).  This design---made from standard printed circuit board (PCB)---provides sample interchangeability without placing much stress on the piezo tube during mounting and dismounting.  The shape of the PCB enables maximum scan range without colliding with the scan head support structure.  Electrical contact is made to a transport device by wirebonding from the copper traces on the PCB to the device.  The layout of the copper traces was chosen to provide ample clearance between the cantilever and device wirebonds.  Wires extend from the PCB and are connected to an interconnect held just above the microscope scan head in the vacuum space.  This intermediate interconnect is used for ease of exchanging samples and to provide wire management.  From this interconnect, wires run to a vacuum feedthrough (Fig. \ref{fig:MicroscopeDesign}(d)).  On the outside of the vacuum can, wires connect the vacuum feedthrough pins to a custom BNC-style breakout box containing the 16 connections.  This breakout box makes circuit reconfigurability simple, as the desired pins can be connected to electrical sources and meters, as well as the FPGA I/Os on its home-built breakout box.  Each pin is controlled by a toggle switch to connect the device lead to either ground or to the connected instrument.  This provides device protection against electrostatic discharge when physically changing the circuit configuration.

\begin{figure}[htp]
	\centering
		\includegraphics[width=\linewidth]{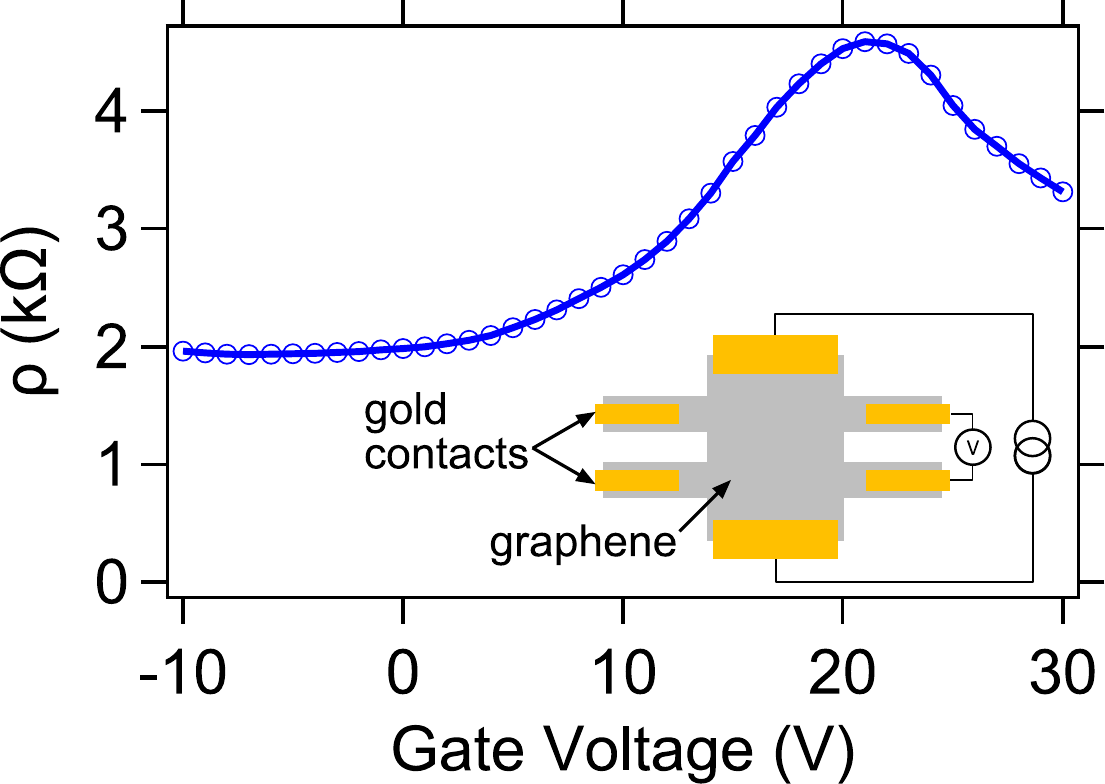}
	\caption{Four point gate-dependent graphene resistivity measurement, acquired in the microscope.  Inset: cartoon of measurement configuration.} 
	\label{fig:4ptGDR}
\end{figure}

An example of a transport measurement obtained with a device mounted in our microscope is shown in Fig. \ref{fig:4ptGDR}.  The device-under-test is a graphene field-effect transistor (FET), patterned in a Hall bar geometry (with multiple pairs of Hall contacts).  Connecting the device in a conventional four-point measurement scheme, we obtain the Dirac-like dependence of the graphene resistivity versus applied back gate voltage \cite{novoselov_electric_2004}.

\subsection{Software}
The microscope control software package contains two sub-programs, one of which is compiled and downloaded for execution on the FPGA, and a second which is executed on the host computer.  The FPGA program provides A/D, D/A, and any time-sensitive, high-speed data processing.  This includes interferometer signal digitization, filtering, cantilever self-excitation, and output of piezo tube control voltages.  The host program serves as the user-interface and performs tasks that are less time-sensitive, such as cantilever signal processing (frequency and amplitude determination), calculation of piezo tube scan voltages (raster and line scan) and other control voltages, various feedback loops, real-time data display, and on-demand data file saving.

At runtime, to obtain digitized data or send control commands, the host program communicates with the FPGA via DMA (direct memory access) transfers that have been established by the FPGA code.  These transfers utilize the MXI data link between the PXI chassis and the PCI bus of the PC.  See Fig. \ref{fig:MicroscopeBlockDiag} for an overview of the host PC and FPGA tasks and connectivity.  The user does not interact directly with the FPGA program at runtime, except through control variables that have been mapped to the host program.  Still, the FPGA program can be modified and recompiled offline if additional functionality or changes to operation are desired.

\subsubsection{Cantilever Self-Excitation}
\label{sec:SelfExcite}
Cantilever self-excitation is a positive feedback method which uses the cantilever's own oscillation to create the signal that feeds back to drive the cantilever.  In this manner, the cantilever is always driven at its self-determined resonance frequency, enabling sensitive frequency shift force detection (see Sec. \ref{sec:FreqDet}).  Conventional amplitude or phase shift detection of forces, where the cantilever is driven at a fixed frequency, experiences a loss of force sensitivity if large forces shift the cantilever response away from the drive frequency by an amount larger than the cantilever bandwidth ($\Delta f \sim f_0/Q$).   There is no risk of this with self-excitation.  

The digitization and processing speed of the FPGA make cantilever self-excitation simple and reliable.  The interferometer signal representing the cantilever position is digitized, delayed by an integer number of FPGA clock-cycles by a $z^{-n}$ Discrete Delay function in the FPGA code, multiplied by a gain factor, and output as a voltage to drive a piezo disc to which the cantilever is mechanically coupled.  (The FPGA clock runs at \SI{40}{\mega\hertz}, enforcing only that the signal be delayed by an integer number of \SI{25}{\nano\second} clock ticks).  When the drive voltage sinusoid is $\pi/2$ out-of-phase with the cantilever position, self-excitation is achieved \cite{albrecht_frequency_1991}.  The user can control the delay time in order to fine-tune the drive phase relative to the cantilever oscillation until the cantilever amplitude is maximized.  The user can also control the drive amplitude, as well as an interrupt time (duration during which the drive is turned off).  In addition, this interrupt time can be controlled by active feedback in order to maintain steady cantilever amplitude during scanning and imaging tasks.

\subsubsection{Cantilever Frequency Detection}
\label{sec:FreqDet}
The frequency of cantilever oscillations is affected by tip-sample interaction and can provide topographic, magnetic, and/or electrostatic information.  The general form of a force acting on a cantilever is given by ${\bf F} = -{\bf \nabla} U$, where $U$ is the potential energy of the cantilever as a function of its position.  The specific form of $U$ will depend on the interaction (topographic, magnetic, electrostatic, etc.).  The force(s) in question will shift the resonance frequency of the cantilever according to 

\begin{equation}
\delta f = -\frac{f_0}{2k} \frac{\partial F(z_0)}{\partial z}
\end{equation}

\noindent for a cantilever with spring constant $k$, resonant frequency $f_0$, and equilibrium position $z_0$.  By detecting this frequency shift as a function of lateral position above the sample, we map the interaction force (although quantitatively extracting the force is difficult).  Using cantilever frequency detection as the imaging modality enables fast and sensitive imaging with high Q cantilevers \cite{albrecht_frequency_1991}.  There is no need to wait for cantilever ``ring-up,'' as in amplitude detection.  The frequency detection algorithm we use works by utilizing the straightforward relationship between a sinusoidal signal and its second derivative; namely that the second derivative is $-\omega^2$ times the original signal. 

\begin{equation}
\frac{\partial^2 \mbox{sin}(\omega t)}{\partial t^2} = -\omega^2 \mbox{sin}(\omega t)
\end{equation}

\noindent Therefore, by calculating the second derivative ($\partial^2 z(t)/\partial t^2$) of the input signal ($z(t)$), and fitting a line to $\partial^2 z(t)/\partial t^2$ vs. $z(t)$, the signal frequency can be determined \cite{obukhov_real_2007}.

This algorithm provides a computationally efficient way to detect the first harmonic of a sinusoidal signal.  It makes use of the entire time record of the oscillatory signal.  This is in contrast to frequency determination by measurement the period between zero-crossings, which discards a majority of the data and is particularly susceptible to noise.  Furthermore, because we are interested in only a single frequency component of the signal, this method is much more precise than performing an FFT on the same time record.  The frequency resolution of an FFT is $1/T$, where $T$ is the length of the entire signal record in seconds.  Therefore, if attempting to obtain \SI{1}{\milli\hertz} frequency resolution, one would need 1000 seconds of data.  This is obviously impractical for scanning applications, where each pixel requires a precise frequency measurement.  We have achieved sub-\SI{1}{\milli\hertz} resolution of a \SI{75}{\kilo\hertz} signal using approximately \SI{3}{\milli\second} of data.

\begin{figure}[htp]
	\centering
		\includegraphics[width=\linewidth]{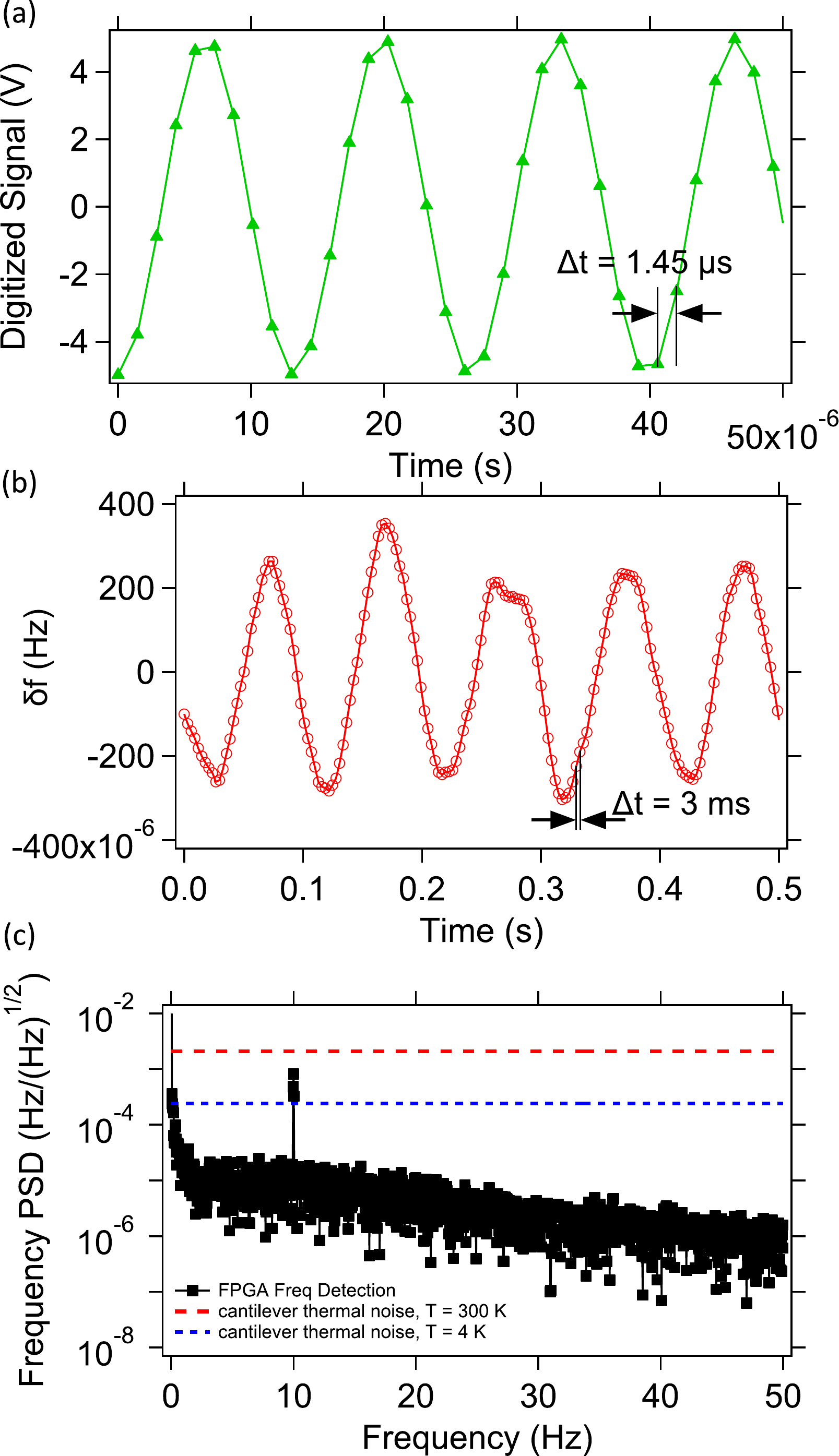}
	\caption{(a) Sinusoidal signal at \SI{75213.833}{\hertz} generated by Stanford Research DS345 function generator, digitized and filtered (bandpass width = 20 Hz), showing the 690 kHz sampling rate.  (b) Frequency vs. time record of the 75,213.833 Hz signal, showing clear frequency modulation: $\sim$1 mHz oscillations at 10 Hz.  Each frequency measurement is calculated from a 2048-data point (3 ms) time record of the sinusoidal signal.  (c) Frequency noise power spectral density (Fourier transform) of the frequency vs. time record, showing the large SNR of the modulation signal at \SI{10}{\hertz}---a 1 mHz peak rising well above the noise floor ($\sim$\SI{10}{\micro\hertz\per\hertz^{1/2}}).  For comparison, the thermal noise limit for a 75 kHz cantilever, oscillation amplitude 20 nm, Q = 10,000 is shown at 300 K and 4 K.}
	\label{fig:FreqPSD}
\end{figure}

To test the limits of our frequency detection algorithm, we generated a frequency modulated sine wave using a Stanford Research DS345 function generator.  The baseline frequency was set to \SI{75213.833}{\hertz} (on the order of many commercially-available cantilevers).  The frequency modulation depth was set to \SI{1}{\milli\hertz} with a modulation frequency of \SI{10}{\hertz}.  The input signal is digitized at \SI{690}{\kilo\hertz} (Fig \ref{fig:FreqPSD}(a)), and 2048 data points are used for each frequency measurement, corresponding to a frequency measurement time of 3 ms, suitable for most scanning applications.  The input signal is filtered by a bandpass of 20 Hz width around the baseline frequency.  Frequency detection results are shown in Fig \ref{fig:FreqPSD}(b,c).  The detected frequency vs. time record (panel (b)), clearly demonstrates the imposed frequency modulation.  In panel (c)---an FFT of (b)---the 1 mHz modulation ``signal'' rises above the noise floor baseline ($\sim$\SI{10}{\micro\hertz\per\hertz^{1/2}}) with an SNR of nearly 100.    

Strictly speaking, the frequency measurement time (\SI{3}{\milli\second}) is shorter than the settling time of the \SI{20}{\hertz} bandpass, resulting in attenuation of the full \SI{1}{\milli\hertz} modulation amplitude.  In practice, we find that the improvement in frequency noise with such a strict filter is justified.  Depending on the measurement, the user can fine-tune the filtering and the scan rate to optimize frequency noise, scan speed, and spatial resolution.  The number of time record data points used for each frequency calculation can also be reduced if necessary and should be determined empirically, depending on the source and magnitude of noise.  Note that presently in our imaging experiments, frequency shift sensitivity is limited not by the frequency detection algorithm, but by the thermal noise of the cantilever $\delta f_{\rm{th}} = F_{\rm{th}} f_0/(2 k x_0)$, where $\delta f_{\rm{th}}$ is the thermally-induced frequency noise (in \SI{}{\hertz\per\hertz^{1/2}}).  The thermal force noise (in \SI{}{\newton\per\hertz^{1/2}}) is calculated as $F_{\rm{th}} = \sqrt{4 k k_B T / 2 \pi f_0 Q}$.  Our system operates in high vacuum ($<$\SI{0.01}{\milli\torr}), at T = \SI{300}{\kelvin}, with a cantilever Q$\approx$10,000 and spring constant $k \sim$\SI{2.8}{\newton\per\meter}.  For comparison purposes, we plot $\delta f_{\rm{th}}$ for this cantilever at \SI{300}{\kelvin} and \SI{4}{\kelvin} in Fig. \ref{fig:FreqPSD}(c) (red and blue dashed lines, respectively).  For ultra-high-Q cantilevers where the thermal noise floor would lie below the FPGA detection noise floor presented in Fig \ref{fig:FreqPSD}(c), the SNR of the input sine wave would need to be improved (for example, by reducing cable noise).  If the input sine wave SNR exceeds 65,536 ($=2^{16}$) the limiting factor becomes the 16-bit digitization noise of the A/D conversion.

The actual code which executes the frequency detection algorithm is shared between the FPGA and host computer.  Because it is memory-intensive to store long arrays of high-precision data, the limited resources of the FPGA are not well-suited to executing the entire algorithm.  Furthermore, a host PC can perform all necessary calculations with double-precision (64-bit) floating point numbers.  This offers immense improvement as compared to the fixed-point and single-precision (32-bit) floating point numbers handled by the FPGA.  As such, the host computer performs the bulk of array manipulations.

\subsubsection{Host Code}
The host code provides the graphical user interface (GUI) with which the user primarily interacts.  Furthermore, much of the instrument control functionality and data processing are provided by the host.  The microscope control software was developed and successfully demonstrated on a standard desktop PC with an Intel Core i7 (2.93GHz) CPU and 8GB of RAM.

\begin{figure}[htp]
	\centering
		\includegraphics[width=\linewidth]{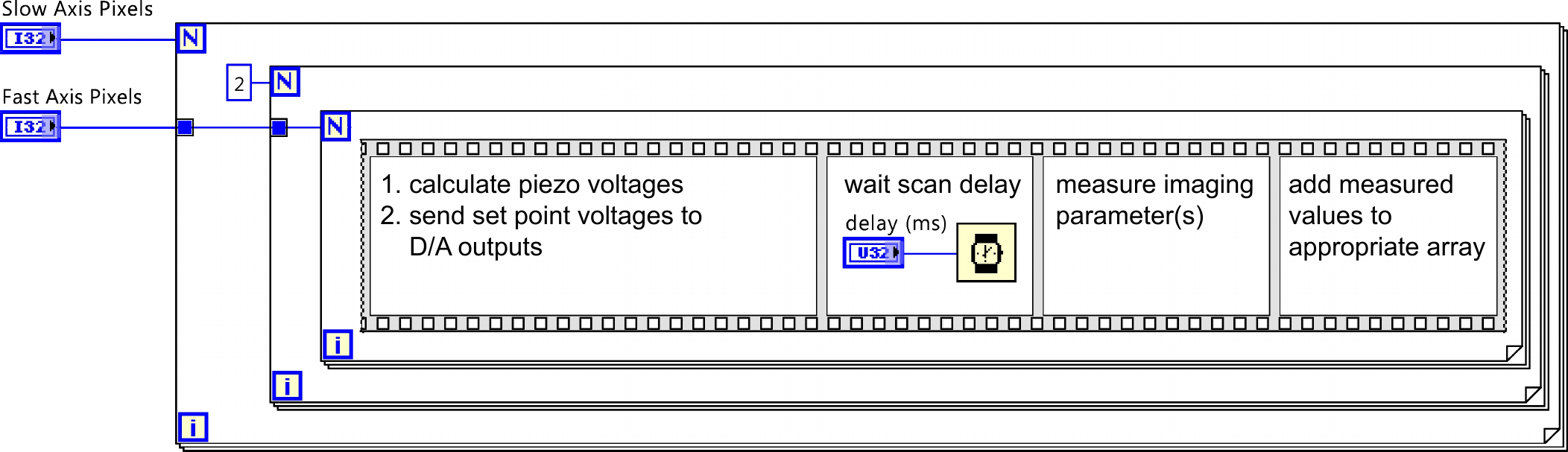}
	\caption{A schematic representation of the LabVIEW scan sequence used for a 2D raster scan.  It is simple to add procedures, data collection and manipulations, etc. to the sequence.} 
	\label{fig:ScanSequence}
\end{figure}

In addition to performing the calculation of the cantilever frequency, the host code is also responsible for calculating the piezo tube voltages necessary for sample approach and raster scanning, performing cantilever amplitude and frequency feedback (if desired), plotting the scanned images (frequency vs. position, for example), and performing on-demand file saving.  To ensure proper sample mapping during imaging tasks, scanning operations must be properly sequenced with measurements of the imaging parameter (e.g. cantilever frequency).  Therefore, LabVIEW queues and notifiers are used extensively for data handling in what is known as a ``producer/consumer'' program design.  This also enables parallelism on the host: data acquired in one loop (producer) can be stored in a queue and accessed by a parallel loop (consumer).  The consumer loop can then perform data processing tasks that would hinder the processing speed of the producer loop if it had been responsible for acquisition \textit{and} processing.  Measurement sequencing is also enforced through the use of interrupts, which are used to notify the host when the FPGA has completed a task (e.g. scanned to a setpoint voltage).  In order to showcase the high-degree of control the user has over the scan sequence, we describe our scanning algorithm below and in Fig. \ref{fig:ScanSequence}.

\begin{enumerate}
\item Mode selection - Raster, Line Scan, Manual Positioning, or Field Scan
\item Read Scan Parameters (start, end, step)
\item Calculate estimated scan time
\item Go to initial position (e.g. set the piezo tube voltage to $(x_0, y_0, z_0)$)
\item Begin scan (see Fig. \ref{fig:ScanSequence})
\end{enumerate}

\begin{figure*}[htb]
	\centering
		\includegraphics[width=\linewidth]{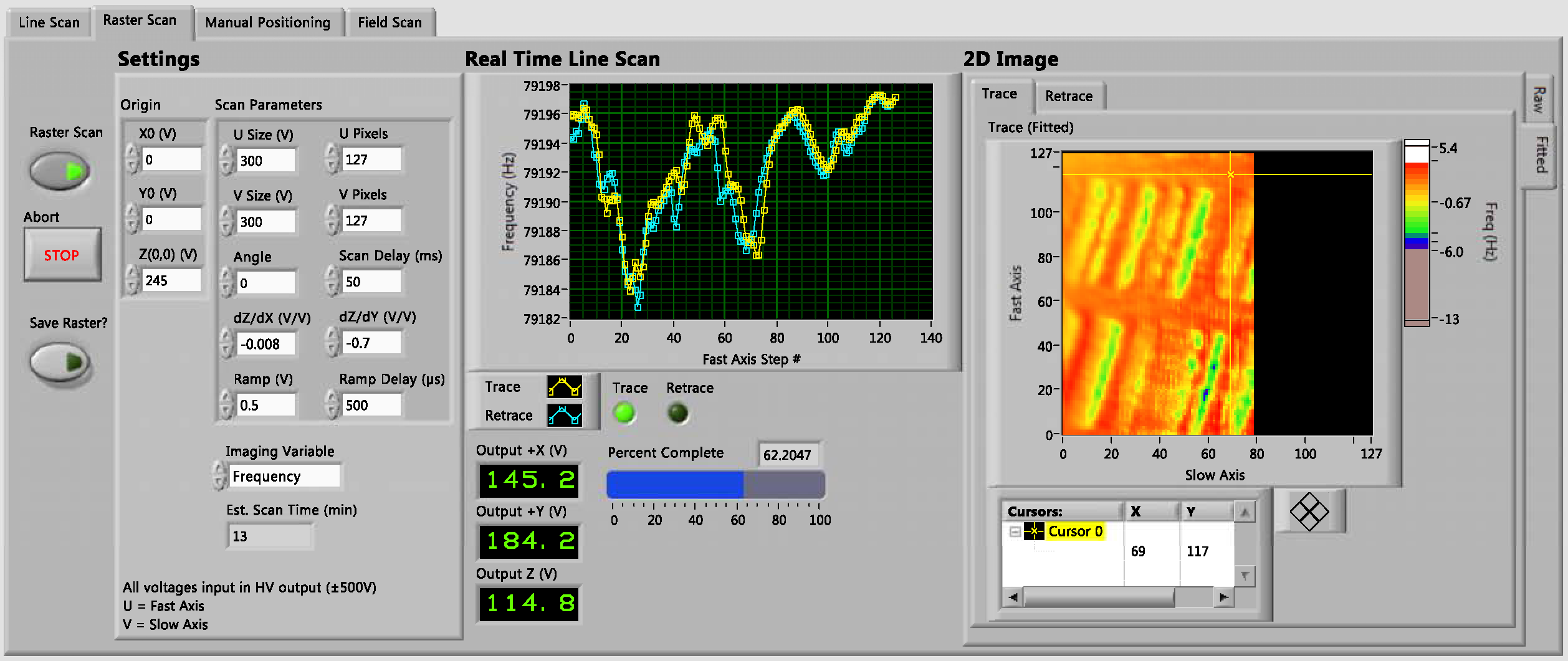}
	\caption{The host GUI showing the Raster Scan controls and indicators.  Visible in three columns from left to right are the scan \textbf{Settings}, 1D \textbf{Real Time Line Scan} (trace and retrace) graphs, and composite \textbf{2D Image}.  The scan settings include the origin; scan size, resolution, angle (rotation of fast and slow axes with respect to the piezo tube x and y axes), and rate; slope correction controls; choice of imaging variable; and estimated scan time indicator.  The real time line scan shows each fast axis trace and retrace scan with real time indicators of the applied voltages to the piezo tube, and a percent complete indicator.  The 2D composite image updates after the completion of each 1D line scan.  The user can view either the raw trace or retrace images, or the ``flattened'' images.  A ``Save Raster?'' control for on-demand file saving of the 2D data is seen below the ``Raster Scan'' (engage) and ``Abort'' buttons.  Also visible are tabs for other scan modes: Line Scan, Manual Positioning (which sets the piezo tube to a specific location (x,y,z)), and a Field Scan option, which sweeps a control voltage for the electromagnet.} 
	\label{fig:RasterScan_FrontPanel}
\end{figure*}

The scanning is handled by a sequence of nested ``for'' loops.  The outside loop is responsible for stepping through the slow axis, the intermediate loop selects the trace or retrace scan, and the innermost loop iterates through the fast axis positions.  The user can easily incorporate acquisition of the desired imaging parameter into this sequence.  After each slow axis step (completion of trace and retrace of the fast axis scan), the results of the trace and retrace line scans are plotted, and the latest scan is added to the composite 2D image.  Because a sample is often mounted with an average tilt, the 2D data is also fed to a real-time plane fitting algorithm in order to flatten the acquired image.  Using LabVIEW's General Linear Fit function, we calculate the best fit plane to the composite 2D image after each line scan is acquired.  This best-fit plane is then subtracted from the raw image data in order to provide a ``flattened'' image.  See Fig. \ref{fig:RasterScan_FrontPanel} for an example of the GUI display during image acquisition.  

\section{Results and Experiments}
\subsection{Standard SPM Operation}
We performed several tests to calibrate the piezo tube motion and demonstrate the performance and capabilities of our microscope.

\subsubsection{Sample approach}
\begin{figure}[htp]
	\centering
		\includegraphics[width=\linewidth]{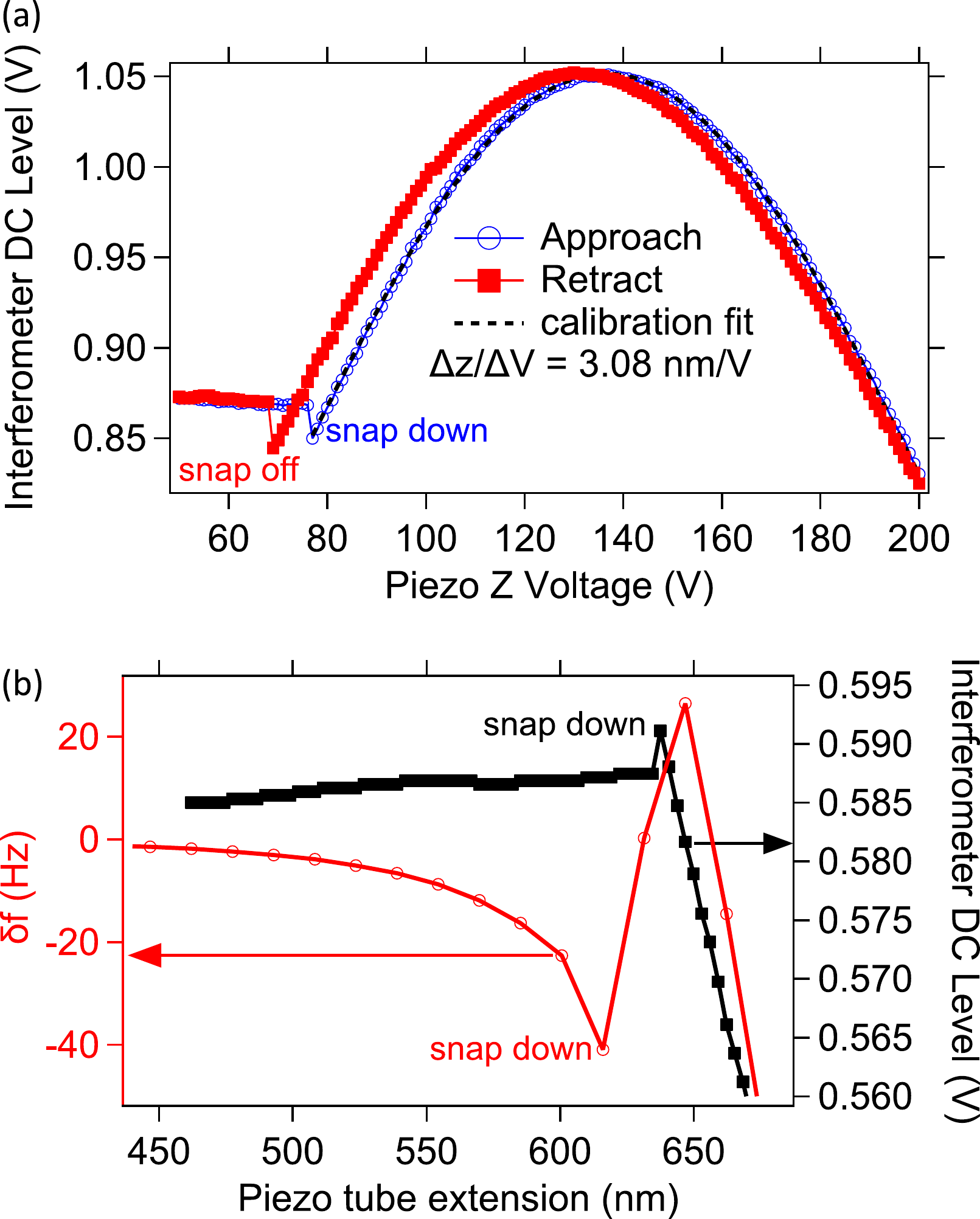}
	\caption{Cantilever-sample approach curves.  (a) Piezo tube voltage-distance calibration.  Approach and retract curves are shown, demonstrating snap-down and off.  The broad sinusoidal oscillation is an interferometer fringe, which can be used for piezo voltage-distance calibration, for which we obtain \SI{3.08}{\nano\meter\per\volt}.  (b) Comparison of frequency-shift (oscillating cantilever) and DC level (undriven cantilever) approach curves.  On a different sample than shown in (a), we acquire frequency shift (red) and interferometer DC level (black) approach traces, taken to establish the sample position, and set a desired scan height.  Note that, due to the oscillations of the cantilever, the frequency shift approach experiences snap down sooner than the DC level approach.  The frequency measurement becomes inaccurate after snap down has occurred.} 
	\label{fig:SampleApproach}
\end{figure}

In order to calibrate the piezo tube extension (in nm per applied volt), we record the interferometer DC level as the sample is raised towards and brought into contact with the cantilever.  Continued extension of the tube causes upward deflection of the cantilever, and a reduction in interferometer cavity length (distance between cantilever and fiber end).  This allows one to observe interferometric oscillations according to $\mbox{sin}((4\pi/\lambda)({\delta z/\delta V})\Delta V)$, where $\lambda$ is the interferometer wavelength (\SI{1550}{\nano\meter}), $\Delta V$ is the applied piezo tube voltage, and $\delta z/\delta V$ is the tube's distance/voltage coefficient \cite{rugar_improved_1989}.  A sinusoidal fit (black dashed line) to the approach curve in Fig. \ref{fig:SampleApproach}(a) yields a calibration coefficient of 3.08 nm/V at room temperature.  

Locating the sample surface is a critical step to setting up an imaging scan.  A coarse tip-sample distance determination is performed as follows: while continuously driving the cantilever, the attocube z motor is stepped upward (towards the cantilever) until the sample comes into gentle contact with the cantilever, causing the oscillations to disappear.  The sample is then retracted by less than \SI{1.5}{\micro\meter}.  We then perform a piezo tube z scan while monitoring either the cantilever frequency (self-excited cantilever, Fig. \ref{fig:SampleApproach}(b), red), or DC level (undriven cantilever, Fig. \ref{fig:SampleApproach}(b), black).  The frequency-measurement approach provides an approximate idea of piezo z voltage necessary to bring the sample into contact with the cantilever.  The DC level approach, because it detects cantilever snap down and snap off, provides a more accurate measurement.  

\subsubsection{Non-contact AFM imaging of a calibration grating}
\label{sec:AFM}
\begin{figure}[htp]
	\centering
		\includegraphics[width=\linewidth]{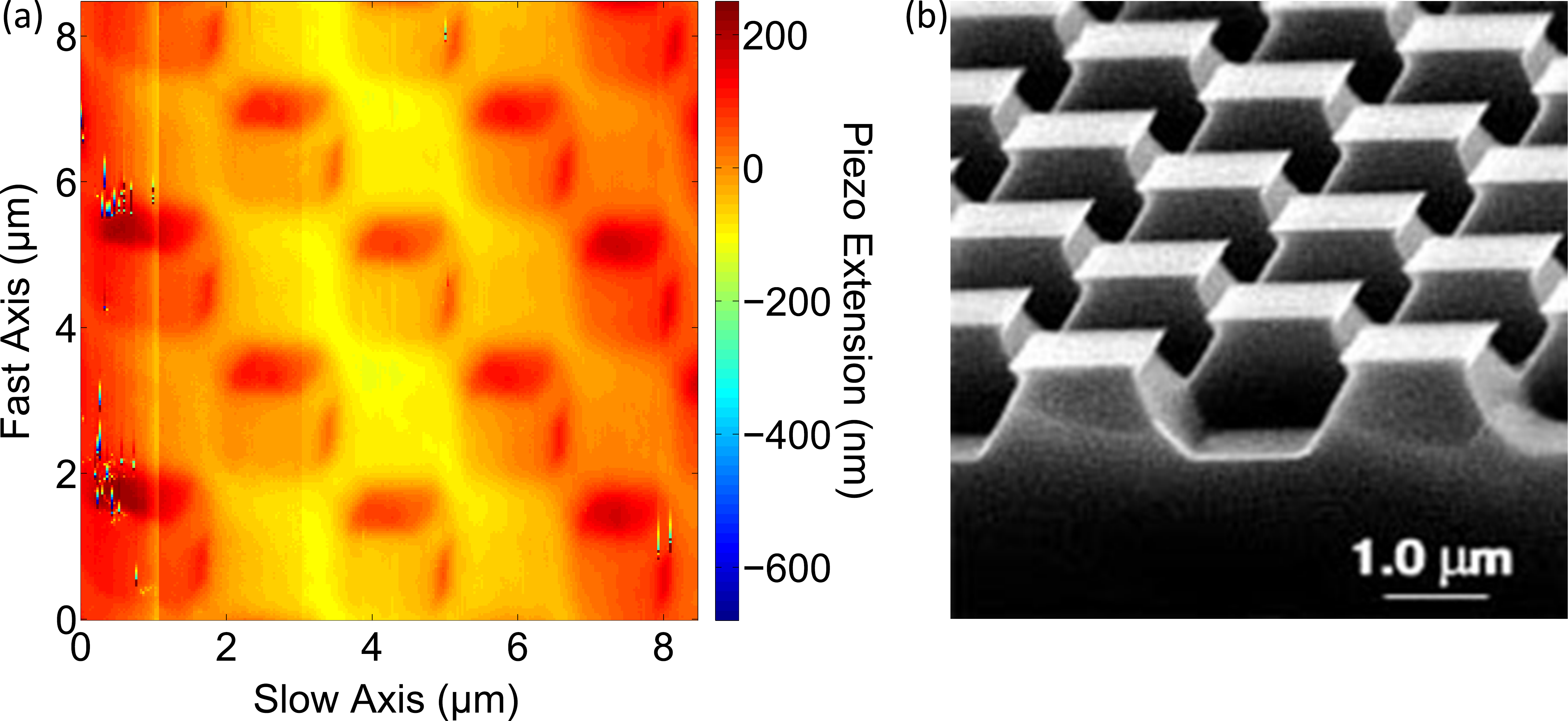}
	\caption{(a) Non-contact AFM image of MikroMasch TGX01 AFM calibration grating (\SI{3}{\micro\meter} pitch), with frequency feedback.  The tip-sample distance is kept constant by adjusting the piezo z voltage until the cantilever frequency reaches the desired setpoint.  This mode is useful for samples with tall features, or when tip-sample contact is not desirable.  The image was acquired at a rate of 50 ms/pixel at a tip-sample distance of 73 nm and oscillation amplitude of 47 nm. (b) SEM image of calibration grating \cite{schaefer_calibration}.} 
	\label{fig:CalibrationGrating}
\end{figure}

We can perform PID control of the cantilever frequency with standard LabVIEW functions, controlling the piezo tube z voltage (extension) with the feedback output.  For samples that interact with the tip purely by Van der Waals forces, this imaging mode corresponds to constant tip-sample spacing.  We imaged a standard AFM calibration grating (MikroMasch \cite{mikromasch} TGX01) to demonstrate this non-contact AFM capability (see Fig. \ref{fig:CalibrationGrating}).  This grating also allows us to calibrate the lateral motion of the piezo scan tube: \SI{28.3}{\nano\meter\per\volt}. 

This image was acquired at a scan speed of 50 ms/pixel.  Since the host PC is responsible for calculating the cantilever frequency, it also handles the frequency feedback.  Iterative loops, as are used in PID, execute much more slowly on a host PC than on the FPGA.  If the cantilever frequency could be calculated on the FPGA (by use of an FPGA card with greater processing capability than the 7851R), feedback could also be executed on the FPGA itself.  This would result in much faster feedback and scanning.

\subsubsection{MFM imaging of a hard disk drive}
\begin{figure}[htp]
	\centering
		\includegraphics[width=\linewidth]{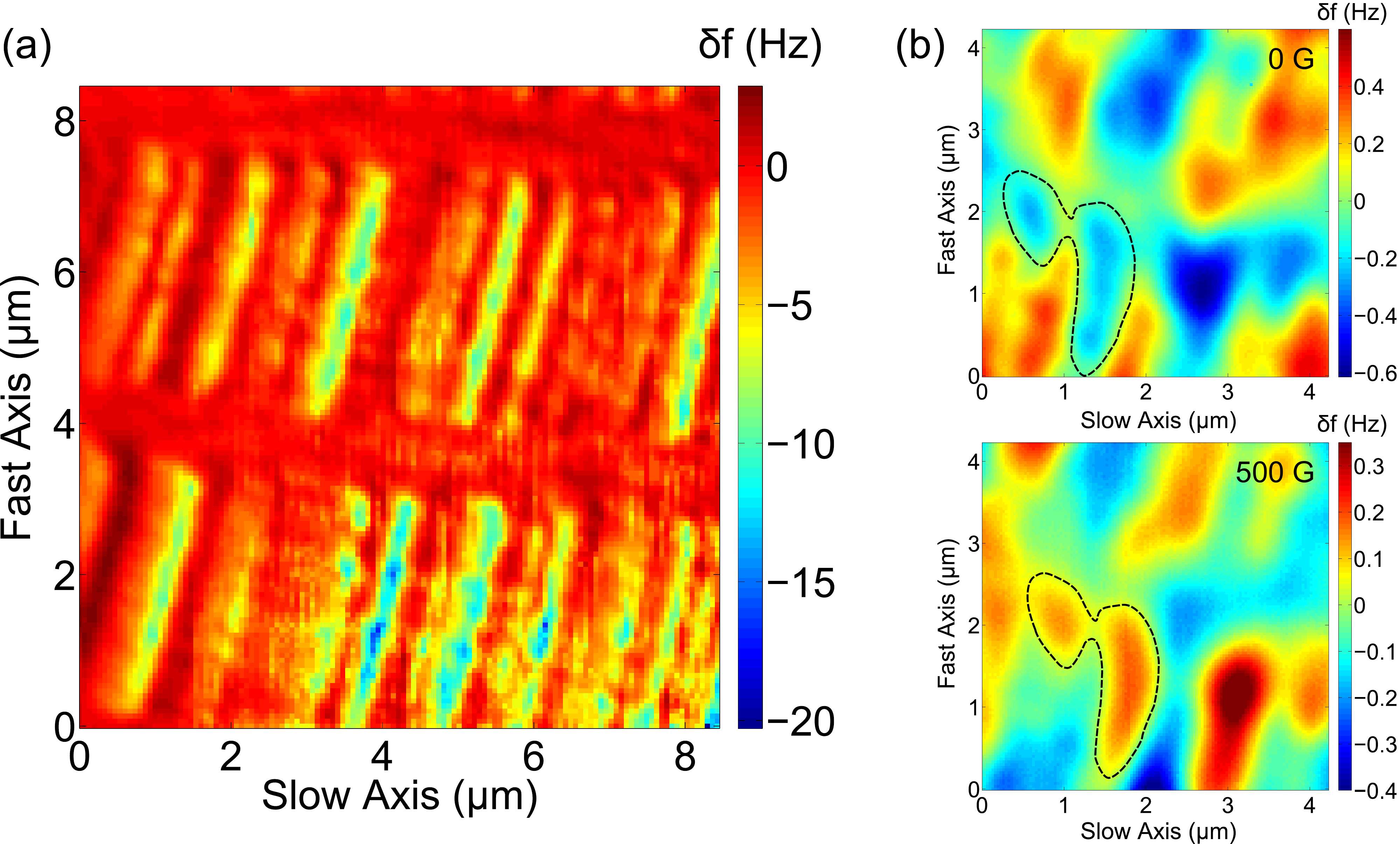}
	\caption{(a) Frequency-shift image of hard-disk drive showing two tracks of data (base cantilever frequency = 79,198 Hz, tip-sample distance = 150 nm, oscillation amplitude = 126 nm). (b) Frequency-shift image of hard-disk drive at zero applied magnetic field (top) and at 500G (bottom).  The magnetization of the cantilever coating has reversed direction (at ~\SI{400}{\gauss}) causing a reversal of the frequency-shift contrast colors.} 
	\label{fig:HardDriveMFM}
\end{figure}

With a magnetically-coated Bruker \cite{brukerAFM} MESP cantilever ($f_0$ = \SI{79198}{\hertz}), we can detect sample magnetization.  Figure \ref{fig:HardDriveMFM}(a) shows a frequency-shift image for a magnetic hard disk drive, showing two tracks of magnetic bit data.  Because magnetic forces can be either attractive or repulsive, the simple frequency feedback mode demonstrated in Sec. \ref{sec:AFM} cannot be used with magnetic samples.  This image was therefore taken at nominally-fixed piezo z voltage.  As a result, some topographic signal may mix into such an image if the tip-sample distance changes during scanning.  A more sophisticated MFM imaging algorithm (e.g. a lift mode) could be incorporated into our software.  Presently, we do incorporate sample tilt correction, which adjusts the piezo z voltage as a linear function of both x and y position.  This tilt can be calibrated by performing sample approaches at three different (x,y) positions.

The microscope was designed to operate in external magnetic fields.  We use a GMW \cite{GMW} 5403 electromagnet atop a rotating stage, in order to provide magnetic field in any in-plane direction (see Fig. \ref{fig:MicroscopeDesign}(d)).  The magnet current is supplied by a Kepco \cite{kepco} 15 V/20 A BOP 15-20M voltage-controlled bipolar operational power supply.  In Fig. \ref{fig:HardDriveMFM}(b), we show a region of the hard drive sample imaged at \SI{0}{\gauss} and \SI{500}{\gauss}.  Cantilever magnetometry \cite{stipe_magnetic_2001} data (not pictured) show that the cantilever magnetization undergoes switching at about \SI{400}{\gauss}.  Such a magnetization reversal will change the sign of the force exerted on the cantilever by an unchanged sample magnetization.  As a result, the color contrast of the frequency shift data in the \SI{0}{\gauss} and \SI{500}{\gauss} images are inverted.  A dashed line is shown as a guide to the eye to show a particular feature where this is evident.

\subsection{In-Operando Imaging: Electrostatic Force Microscopy}
\label{sec:EFM}
We acquired scanned electrostatic force microscopy (EFM) images of an electrically-biased graphene field effect transistor (CVD graphene on SiO2(300nm)/n-Si) in order to demonstrate the ability to perform integrated scanning and current-voltage measurements.  In an EFM measurement, the cantilever frequency is shifted by the capacitive interaction between tip and sample \cite{girard_electrostatic_2001}. 

\begin{equation}
\delta f = - \frac{1}{2} \left( \frac{d^2 C}{dz^2} \right) V^2
\end{equation}

\noindent where $C$ is the capacitance of the tip-sample system, and $V$ is the potential difference between tip and surface.  

\begin{figure*}[htb]
	\centering
		\includegraphics[width=\linewidth]{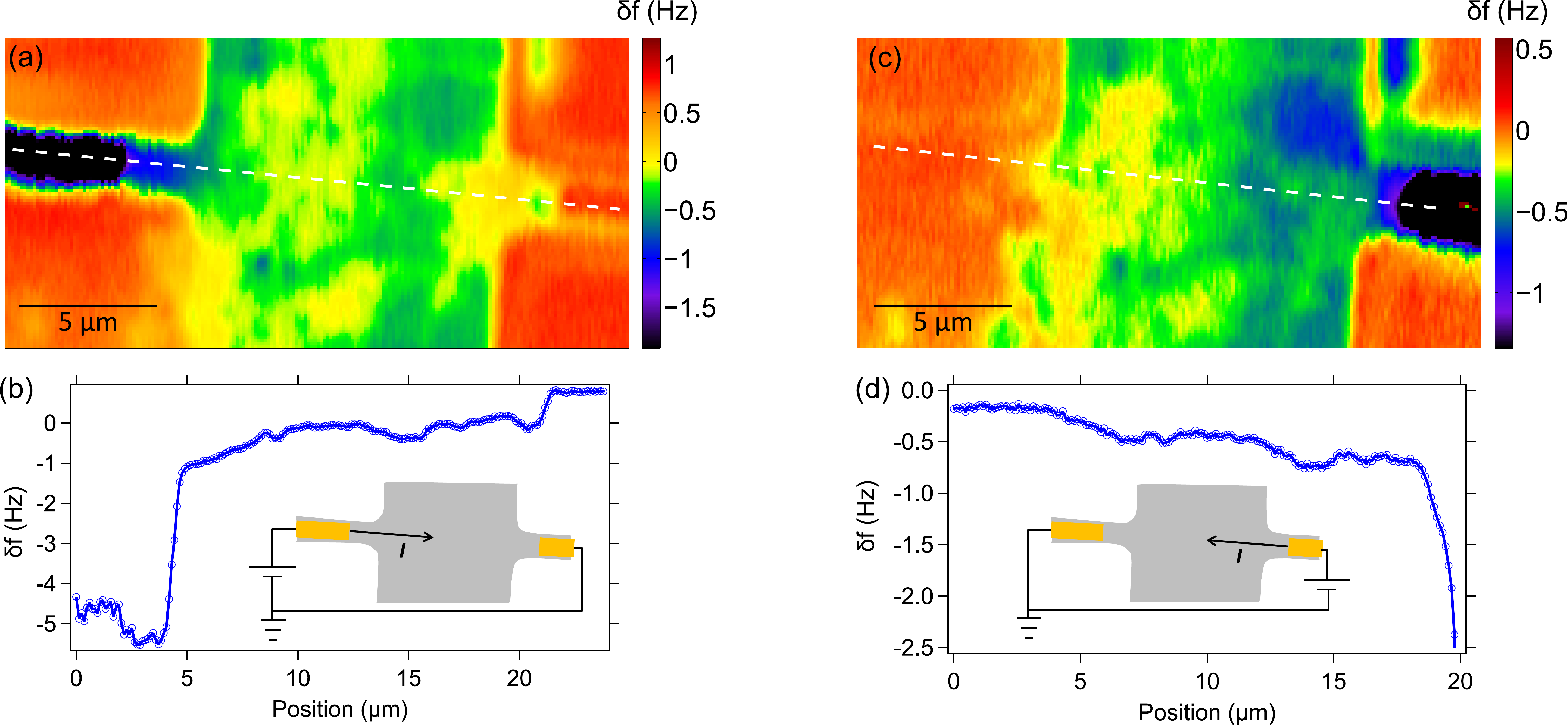}
	\caption{EFM images of a biased graphene hall cross. (a) \SI{100}{\micro\ampere} applied with left contact at $V = +V_0$, right contact at ground. (c) \SI{100}{\micro\ampere} applied with right contact at $V = +V_0$, left contact at ground.  Panel (b) shows a the line-cut indicated by white dashes in (a), while (d) represents the line-cut from (c).} 
	\label{fig:SKFPM}
\end{figure*}

In Fig. \ref{fig:SKFPM} we show the measured cantilever frequency shift for two configurations of the graphene hall cross: electrical current flowing left to right, or right to left.  The cantilever is sensitive only to voltage differences between itself and the sample.  The cantilever (which has a conductive chromium coating) is grounded, so the grounded electrical contact causes a negligible frequency shift.  By contrast, the biased electrical contact shows a dramatic 2 Hz frequency shift relative to the cantilever's natural frequency.  By incorporating a closed-loop Kelvin probe microscopy KPM technique, where the cantilever bias $V_{\rm{probe}}$ is adjusted in order to null the sample's contact potential difference $V_{\rm{CPD}}$ (instead of using a grounded cantilever, as presently), the sample's surface potential (in units of volts) could be directly measured.  The technique can also be used to characterize the quality of contact to graphene \cite{yu_tuning_2009}.  Even without a KPM controller, Fig. \ref{fig:SKFPM} makes evident the voltage drops due to contact resistance between the gold electrodes and the graphene.  Voltage drops in the graphene itself are also visible, particularly in the narrow regions, but appear smaller than those from the contact resistance due to the $V^2$ dependence of $\delta f$.  Three-point current-voltage measurements of the gold/graphene contacts found contact resistances of \SI{11.5}{\kilo\ohm} and \SI{10.3}{\kilo\ohm} for the left and right contacts, respectively.  The total two-point resistance between the pair was found to be 36.6 k$\Omega$, leaving the graphene channel with a \SI{14.8}{\kilo\ohm} resistance.  Because the graphene channel extends far above and below the image ($\sim$\SI{100}{\micro\meter}), current spreading is also observed.  The inhomogeneity in local voltage requires further investigation.  

\begin{figure}[htb]
	\centering
		\includegraphics[width=\linewidth]{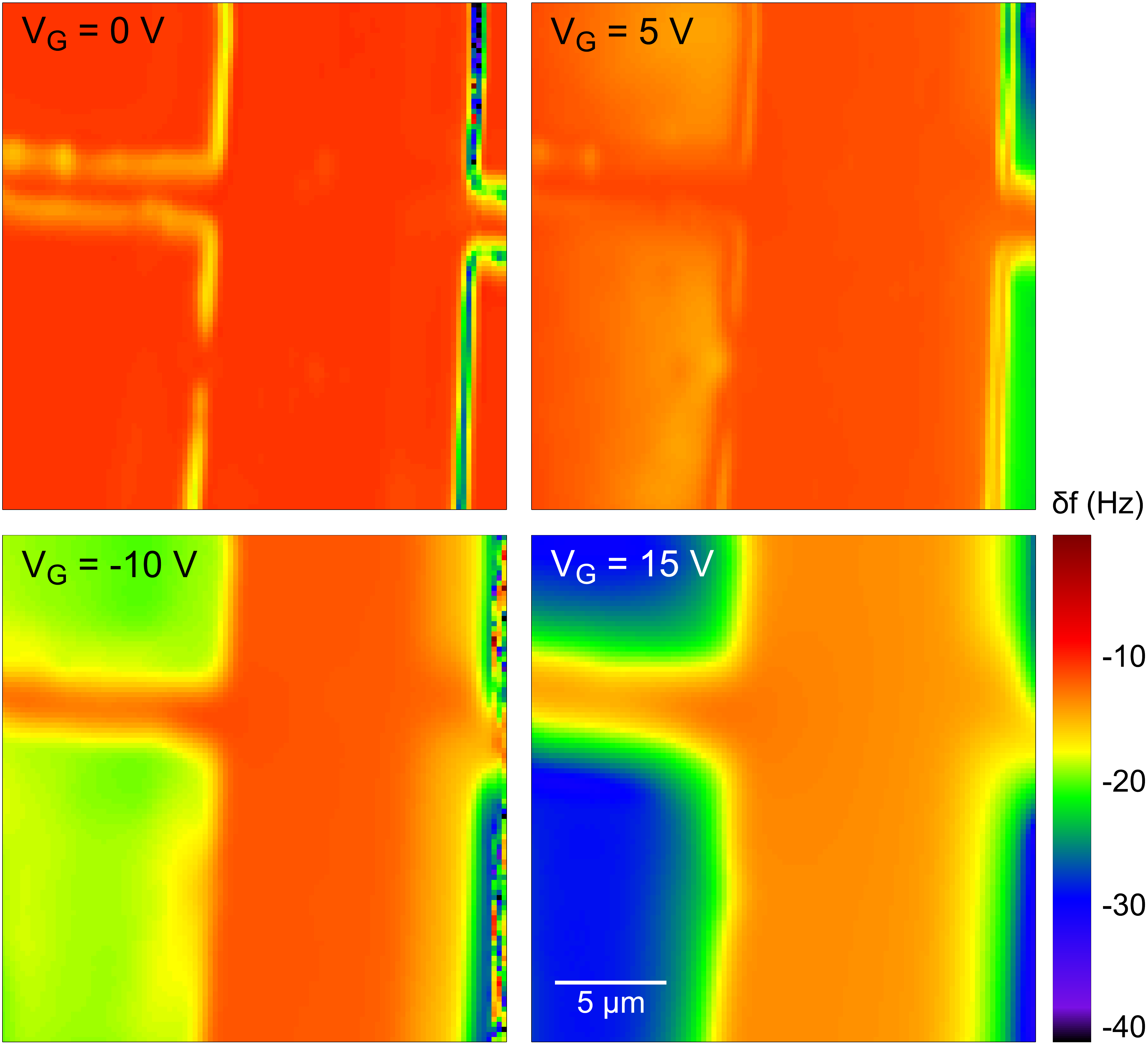}
	\caption{EFM images of the same graphene hall cross as Fig. \ref{fig:SKFPM}, under varying gate bias. The force on the cantilever is sensitive only to $V^2$, and insensitive to the sign of the gate bias.  When held above the graphene (and gold electrodes), the cantilever is shielded from the applied back gate, and experiences relatively little frequency shift.  This technique could be used to easily locate a device, since the electrostatic force is fairly long-ranged \cite{li_self-navigation_2011}.  All images, with the exception of $V_G$ = \SI{15}{\volt}, were taken at a nominal scan height of 340 nm.  The cantilever was retracted to a height of 970 nm in order to acquire the image with $V_G$ = \SI{15}{\volt}, due to the increased electrostatic force causing tip-sample contact.} 
	\label{fig:GateDependence}
\end{figure}

We have also acquired images under varying back gate conditions, as shown in Fig. \ref{fig:GateDependence}.  The gate potential exerts a force on the cantilever, shifting its frequency.  The image contrast is provided by shielding of the applied back gate by the grounded device (graphene and gold contacts), above which the cantilever experiences almost no frequency shift.  The force on the cantilever does not depend on the sign of the voltage difference (since it is proportional to $V^2$).  For example, an attractive force (negative frequency shift) is evident above the unshielded gate for both \SI{-10}{\volt} and +\SI{15}{\volt}.  Because of the long range of electrostatic forces as compared to topographic forces, these images suggest a means for helping to locate a device \cite{li_self-navigation_2011}.

\section{Further Improvements}
Because of the flexibility of the software and hardware we are using, it would be straightforward to expand the imaging and measurement modes of this instrument beyond cantilever force detection.  For example, using a biased conducting tip, the device conductance could be monitored as a function of tip position (scanned gate imaging) \cite{topinka_imaging_2000, topinka_coherent_2001, berezovsky_imaging_2010}.  Alternatively, the Hall voltage of a device such as that used in Fig. \ref{fig:SKFPM} could be monitored in order to quantify the stray magnetic field of an MFM tip \cite{panchal_magnetic_2013}.  We plan to use the microscope to create a spin map of a lateral spin valve by monitoring the effect of a magnetic cantilever tip on the device's non-local voltage, as in scanned spin-precession imaging \cite{bhallamudi_imaging_2012, bhallamudi_experimental_2013}.  Additional measurement protocols could be added in order to perform sensitive measurements.  For example, the cantilever oscillations could be used as a lock-in reference, while monitoring the device voltage.  Again, the modularity of the PXI chassis, the reprogrammability of the FPGA and control software, and the ease of interfacing LabVIEW with external instrumentation will make such measurements possible and relatively easy to implement.

\section{Conclusion}
Scanned probe microscopy plays a central role in micro- and nano-scale characterization of samples and devices.  Combining scanned probes with operational devices is not a capability well-supported by commercial SPM solutions.  The variety of transport measurements and effects that can be studied puts a premium on reconfigurability.  LabVIEW places measurement control in the hands of the experimenter, enabling nearly endless combinations of scanning and transport measurements and protocols.  FPGA-based operation is a perfect fit for imaging tasks because of its speed and deterministic hardware execution.  As such, the combination of FPGA and LabVIEW enables low-cost, versatile in-operando SPM solutions.  Using this platform, we have demonstrated fast and accurate frequency shift detection and imaging, straightforward and reliable incorporation of transport measurements, and the flexibility to pursue unique and innovative measurement schemes for investigation of electronic and spintronic systems.

\section{Acknowledgments}
This research was supported by funding from the Center for Emergent Materials: an NSF MRSEC, Award Number DMR-1420451 and by the Army Research Office, ARO Award No. W911NF-12-1-0587.  Technical support was provided by the NanoSystems Laboratory at OSU.


\bibliographystyle{unsrtnat}
\bibliography{References}

\end{document}